\documentclass[%
reprint,
showpacs,
english,amsmath,amssymb,aps,pre]{revtex4-1}

\usepackage{graphicx}
\usepackage{epsfig}
\usepackage{dcolumn}
\usepackage{bm}
\usepackage{gensymb} 

\usepackage[T1]{fontenc} 
\usepackage{lmodern}
\usepackage{color}

\begin{document}

\title{Stochastic Rotation Dynamics for Nematic Liquid Crystals}
\author{Kuang-Wu Lee}
\email{jeff.lee@ds.mpg.de}
\author{Marco G. Mazza}
\email{marco.mazza@ds.mpg.de}
\affiliation{Max Planck Institute for Dynamics and Self-Organization, Am Fa{\ss}berg 17,
37077 G\"ottingen, Germany}

\date{\today}

\begin{abstract}
We introduce a new mesoscopic model for nematic liquid crystals (LCs). We extend the particle-based stochastic rotation dynamics method, which reproduces the Navier-Stokes equation, to anisotropic fluids by including a simplified Ericksen-Leslie formulation of nematodynamics.  We verify the applicability of this hybrid model by studying the equilibrium isotropic-nematic phase transition and nonequilibrium problems, such as the dynamics of topological defects, and the rheology of sheared LCs. Our simulation results show that this hybrid model captures many essential aspects of LC physics at the mesoscopic scale, while preserving microscopic thermal fluctuations.

\end{abstract}
\pacs{61.30.-v, 64.70.M-, 83.80.Xz}
\maketitle

\section{Introduction}
\label{sec:Introduction}

Liquid crystals (LCs) possess anisotropic interactions because of their molecular shape. 
For example, molecules with a rod-like rigid core tend to align parallel to each other and form a mesophase called nematic; molecules with a disk-like rigid core form discotic phases. In both cases the rigidity is generated by different combinations of aromatic rings~\citep{deGennes:1993}.
Macroscopically, this anisotropy leads to a series of phase transitions that break rotational and translational symmetries in a step-wise fashion. 
Because of their capacity to reorient, also in response to external fields, LCs are used in a wide range of applications: from the ubiquitous electronic displays, to microlasers \cite{humarNatPhot2009,humar-OE-2010,peddireddy-OE-2013} and lubricants \cite{Amann-2013}; but they are also rising to an important role in  biomedical sciences and applications \cite{stewart2003liquid}, and in our comprehension of morphogenesis and evolution of living organisms \cite{stewart2004liquid}.

Hydrodynamic flow can also couple with the local preferential direction (director) established in nematic LCs~\cite{Stephen-1974}.
Recently, Sengupta \emph{et al}.~have extended microfluidic applications to anisotropic fluids
and have found surprising topologies of the orientational field \cite{Sengupta:2013,Sengupta-cyl-SM-13}, and explored fluid and colloidal transport~\cite{Sengupta-SM-13,Sengupta-LC-14,Sengupta-IJMS-13}. These effects originate from the intimate connection between LC rheological properties \cite{Parodi:1970} and their local alignment, which can be easily controlled.
%
%
However, the complex interplay of confinement to a mesoscopic scale, i.e. of the order of $\mu$m, surface interactions, hydrodynamic flow and generation of topological defects still poses formidable challenges to both theoretical and experimental investigations.




The theoretical description of LCs based on static continuum theory started in the early 1920s with the work of Oseen \cite{Oseen}, Zocher \cite{Zocher}, and Frank \cite{Frank:1958}. The earliest dynamics theory of LCs can be dated back to 1931 by Anzelius \cite{Anzelius}. In the 1960s Ericksen and Leslie developed a hydrodynamics theory~\cite{ericksen1959,Ericksen:1960,Ericksen:1961,leslie1966,Leslie:1968}
based on the LC's velocity field $\vec v(\vec r)$ and a unit vector describing the local director $\vec d(\vec r)$. The nematodynamic equations of the Ericksen-Leslie model are widely used but rest on the assumption that the nematic order parameter is a constant and the nematic LC is uniaxial, and 
therefore they cannot describe physical situations where there is a strong variation of the the nematic order, such as the isotropic-nematic phase transition and the dynamic of topological defects. 
For cases where there is a strong variation of the nematic order a tensorial description is necessary, such as the Beris--Edwards formulation \cite{beris1994thermodynamics} or the Qian--Sheng formulation~\cite{Qian-Sheng-PRE}. These two approaches differ in the form of the elastic free energy, that is, the former considers the elastic free energy in the one-constant approximation, while the latter allows for two different elastic constants. We are not aware of any tensorial description of nematodynamics in terms of all three elastic constants, \emph{i.e.} splay, twist and bend. 

Microscopic models have in general the advantage of providing detailed
dynamics at small spatial and temporal scales. Molecular dynamics simulations are well suited for this task. However, physical phenomena at the mesoscopic scale are still so computational demanding that they are out of the reach of atomistic simulations. This predicament can be ameliorated by adopting a coarse-grained description of the molecular degrees of freedom that effectively includes hydrodynamic modes. 
One widely-used method is the lattice Boltzmann scheme that simulates the evolution of the Boltzmann equation for a simple fluid on a regular lattice by a series of collision and propagation steps for probability density functions defined on the sites of a lattice \cite{McnamaraZanettiPRL1988}. This scheme has been generalized to LCs with successful results \cite{Denniston-EPL-2000,CareJPCM2000,CarePRE2003}. However, lattice Boltzmann schemes have the limitation that they do not include thermal fluctuations.  A different approach that emphasize the particle aspect is the dissipative particle dynamics (DPD). This is an off-lattice, particle-based method, where particles represents fluid elements and are subject to pair-wise additive forces, which conserve momentum locally and thus generate the Navier-Stokes hydrodynamics for simple fluids. The DPD scheme has also been extended to LCs \cite{AlSunaidi2004,LevineJCP2005}. Whereas this method has been shown to reproduce equilibrium phase diagrams of LCs~\cite{AlSunaidi2004,LevineJCP2005}, we are not aware of any attempt at reproducing nematodynamic behavior, though the original version of DPD  does reproduce the correct hydrodynamics of simple fluids \cite{HoogerbruggeEPL1992}.

More recently, Malevanets and Kapral \cite{Malevanets:1999} introduced the stochastic rotation dynamics (SRD) model. This is also an off-lattice, particle-based model where each particle represents a small parcel of fluid. The fluid evolves through a series of collisions and streaming steps that exactly conserve mass, linear momentum and energy, and additionally respect Galilean symmetry. Thus, the correct hydrodynamic modes are generated. Because SRD is a particle-based method, fluctuations are naturally present. The SRD method has been applied to a variety of systems, from colloids \cite{PAddingPRL2004}
to polymers~\cite{malevanets2000dynamics,WinklerJPCM2004}, from the modeling of the solvent carrying hydrodynamic interactions in vescicle self-assembly~\cite{NoguchiJCP2006} to the flow-induced shape transition in red blood cells~\cite{NoguchiPNAS2005}. In general the SRD model is useful whenever both thermal fluctuations and hydrodynamic modes are physically important.

In the present work we extend the SRD model to anisotropic fluids, namely LCs. This is primarily done by 
giving orientational degrees of freedom to the SRD particles which obey the (simplified) equations of nematodynamics in the formulation of Ericksen-Leslie. For the sake of simplicity we restrict ourselves to methods and results in two dimensions (2D).
We show below that the SRD is amenable of studying both the equilibrium and nonequilibrium behavior of LCs. 
This new extension of the SRD model opens the door to the investigation of a wealth of LC phenomena occurring at the mesoscale. 



This work is organized as follows. In Sec.~\ref{sec:SRD_LC} we describe the theoretical background of the equations used in the present model, and the details of the numerical implementation.  We validate this model for nematic LCs by considering three study cases in Sec.~\ref{sec:applications}: i) the  
isotropic-nematic phase transition; ii) the production and annihilation of topological charges, and
iii) LC rheology under shear flow. Finally, we summarize and discuss our results in Sec.~\ref{sec:Discussions}.

\section{The Model}
\label{sec:SRD_LC}

\subsection{Theoretical Background}

We start by considering the standard SRD model that  has so far been used to describe fluids with isotropic interactions.
The system is composed of $N$ particles of mass $m_i$ with positions $\vec r_i(t)$ and velocities $\vec v_i(t)$, where $i\in[1,N]$.  The evolution in time $t$ proceeds through a series of two steps:\\ 
(i) the free-streaming step
\begin{equation}\label{eq:SRD-pos}
\vec r_i(t+\delta t)=\vec r_i(t)+\vec v_i(t)\delta t\,,
\end{equation}
where all positions are updated;\\
(ii) the collision step 
\begin{equation}\label{eq:SRD-vel}
\vec v_i(t+\delta t)=\vec u_{C_i}(t)+\bm{\mathrm{R}}\left[\vec v_i(t)-\vec u_{C_i}(t)\right]
\end{equation}
where all velocities are updated by rotating the fluctuating part of the velocity with respect to the center of mass velocity
\begin{align}\label{eq:SRD-cm}
\vec u_{C_i}(t)=\frac{1}{M_{C_i}}\sum_{i=1}^{\mathcal{N}_{C_i}} m_i\vec v_i\,, \quad &
M_{C_i}\equiv \sum\limits_{i=1}^{\mathcal{N}_{C_i}} m_i
\end{align}
In Eq.~(\ref{eq:SRD-vel}-\ref{eq:SRD-cm}) the calculations are performed in a cell-wise fashion, that is, the system is divided with a regular grid, and $\vec u_{C_i}$ is computed from the $\mathcal{N}_{C_i}$ particles within the cell $C_i$ to which particle $i$ belongs.

The rotation (or collision) matrix $\bm{\mathrm{R}}$ is orthogonal, $\bm{\mathrm{R}}^{-1}=\bm{\mathrm{R}}^\mathrm{T}$, where the superscript $\mathrm{T}$ denotes transposition. It rotates, independently in each cell, the fluctuating part of the velocity by an angle $\alpha$ about an arbitrary axis. Because of its action on the fluctuating part of the velocity the collision rule conserves momentum; because of its orthogonality the energy is also conserved. From these facts follows that the fluid obeys the Navier--Stokes equations. Malevanets and Kapral showed \cite{Malevanets:1999} that if  $\bm{\mathrm{R}}$ satisfies detailed balance then the fluid approaches a Maxwell--Boltzmann distribution and obeys the $H$-theorem. In practice there are many ways to choose $\bm{\mathrm{R}}$ so that all the requirements are met. In 2D $\bm{\mathrm{R}}$ can only rotate the velocities by an angle $\pm\alpha$ with equal probabilities; in 3D it is common, \emph{e.g.}, to choose a random axis and rotate around it of a fixed angle.
To ensure Galilean invariance the grid must be shifted by a random amount at each time-step~\cite{IhlePRE2001} so that artificial correlations among particles do not build up due to repeated collisions with the same neighbors.


We now come to the extension to nematic LCs. The particle's degrees of freedom must be augmented by a unit vector describing its orientation $\vec d_{i}$, $\|\vec d_i\|^2=1$. Two particles separated by a distance $\|\vec r_i-\vec r_j\|\leqslant \epsilon$ interact through the Lebwohl-Lasher potential \cite{Lebwohl-PRA-1972}
\begin{equation}
\nonumber
U_{i} = -\sum_{\left\langle i,j\right\rangle} (\vec d_{i} \cdot \vec d_{j})^2.
\end{equation}
where $\left\langle i,j\right\rangle$ indicates that $i$ and $j$ are neighbors. 

Lin \emph{et al.} \cite{Lin-1989,Lin-1995} proposed a simplified version of the Ericksen-Leslie equations. These are nonparabolic, dissipative equations that describe the flow of nematics in the one-constant approximation. Although they represent a drastic simplification of the original Ericksen-Leslie model, they retain the essential mathematical features, and they obey an energy law similar to the one used in the Ericksen-Leslie model \cite{Lin-1995}.

We propose the following hybrid model of SRD for anisotropic fluids
\begin{gather}
\frac{\partial\vec v}{\partial t}+ \vec v\cdot \nabla \vec v =\nabla\cdot(\nu\nabla\vec v)-\nabla P/\rho
-\lambda\nabla \cdot \bm{\pi}
\label{Navier-Stokes}
\\
\frac{\partial\vec d}{\partial t}+ \vec v\cdot \nabla \vec d
- \vec d \cdot \nabla\vec v = \gamma_{_\mathrm{EL}} \nabla^2 \vec d - \gamma f(\vec d) + \vec \xi(t)
\label{EL-director}
\end{gather}
where $\rho$ is the density, $P$ the pressure, $\nu=\eta/\rho$ the kinematic viscosity, and  $\bm{\pi}=(\nabla {\vec d}^{\,\,\mathrm{T}} \cdot\nabla \vec d)$ the Ericksen-Leslie stress tensor, that is the tensor whose $(\alpha\beta)$ component is $\partial\vec d/\partial r_\alpha \cdot \partial\vec d/\partial r_\beta$. In Eq.~\eqref{EL-director},
$\gamma_{_\mathrm{EL}}$  is  the elastic relaxation constant, the term $f(\vec d_{i}) = \partial U_{i}/ \partial \vec d_{i}$ is the molecular field inducing nematic ordering, $\gamma$ its strength, and $\vec{\xi}(t)$ is a Gaussian white noise for
the director's angular velocity, with $\left\langle {\xi}_\alpha \right\rangle=0$ and $\left\langle  \xi_\alpha(t)\xi_\beta(t')\right\rangle=2k_\mathrm{B}T\gamma\delta_{\alpha\beta}\delta(t-t')$, where $k_\mathrm{B}$ is Boltzmann's constant, $T$ is the temperature, $\delta_{\alpha\beta}$ is a Kronecker delta, and $\delta(t)$ is Dirac's delta distribution.


The  Ericksen-Leslie stress tensor $\bm{\pi}$ represents the feedback of the director field $\vec d (\vec r,t)$ to the bulk flow of molecules. In this formulation, $\bm{\pi}$  is directly responsible for the non-Newtonian behavior of the LC flow. Because of their molecular anisotropy the viscosity of a LC fluid does not depend solely on the shear stress $\tau = \eta\partial v/\partial x$, but the director field $\vec d (\vec r,t)$ also participates in the viscosity generation. This means that the velocity field can couple to the director and reorient it, and, also, that a reorientation of the director may generate a flow, usually called \emph{backflow}.

\subsection{Numerical Implementation}

We now describe the numerical implementation of our hybrid model for LCs. We will make a distinction between quantities computed on a particle level, such as $\vec v_i$, and on a cell level, such as $\vec u_{C_i}$. 
We note that the standard SRD steps described in Eqs.~(\ref{eq:SRD-pos}-\ref{eq:SRD-vel}) recover the Navier--Stokes equation, that is Eq.~\eqref{Navier-Stokes} without the last term. Also, Eq.~(\ref{EL-director}) applies in the Eulerian picture (lab-frame) to the  fluid parcel moving along the streamline. 
Since in simulations the director of the $i$th particle, $\vec d_{i}$, moves along the flow, the convective term $\vec v \cdot \nabla\vec d$ is absorbed in this Lagrangian picture (comoving frame).


The SRD algorithm for LCs consists of the following steps: \\
(i) free streaming
\begin{equation}\label{eq:LC-SRD-pos}
\vec r_i(t+\delta t)=\vec r_i(t)+\vec v_i(t)\delta t\,,
\end{equation}
this is identical to the standard SRD step; \\
(ii) cell-wise calculations, that is, 
the particles are grouped in different cells and $\vec u_{C_i}$, $\vec d_{C_i}$,  gradients such as $\nabla \vec d$ and  $\nabla \vec v$, and the Ericksen-Leslie elasticity tensor $\bm{\pi}$ are also calculated;\\
(iii) LC alignment
\begin{multline}
\label{eq:LC-SRD-dir}
\vec d_i(t+\delta t)=\vec d_i(t)+\left[\vec d \cdot \nabla\vec v +\gamma_{_\mathrm{EL}} \nabla^2 \vec d \right.\\
- \left.\gamma f(\vec d) + \vec \xi(t) \right]\delta t
\end{multline}
where Eq.~\eqref{EL-director} is implemented.\\
(iv) Collisions and backflow. The cell-wise, center of mass velocity is calculated as in Eq.~\eqref{eq:SRD-cm}, and the contribution from the Ericksen-Leslie tensor is then added
\begin{gather}
\vec{u}^{\,'}_{C_i}(t)=\vec{u}^{\,}_{C_i}(t)+\lambda\nabla \cdot \bm{\pi}_{C_i}\\
\label{eq:LC-SRD-vel}
\vec v_i(t+\delta t)=\vec u^{\,'}_{C_i}(t)+\beta_\mathrm{th}\bm{\mathrm{R}}\left[\vec v_i(t)-\vec u_{C_i}(t)\right]
\end{gather}
We consider here only a 2D system, thus the matrix $\bm{\mathrm{R}}$  rotates the particles' thermal velocities around one axis, conventionally denoted as the $z$-axis, either clockwise or counter-clockwise by a fixed angle $\alpha$ stochastically. The parameter $\beta_\mathrm{th}$ is the thermostat scaling factor whose role will be explained below.

In general, SRD does not conserve angular momentum. 
To impose the conservation of the fluid's angular momentum in 2D the angle $\alpha$ must be chosen \cite{Ryder2005,Gompper:2008} such that
\begin{align}
\sin\alpha=-\frac{2AB}{A^2+B^2}\,, \quad & \cos\alpha=\frac{A^2-B^2}{A^2+B^2}
\end{align}
where
\begin{subequations}
\begin{align}
A&=\sum_{i=1}^{\mathcal{N}_{C_i}}\left[\vec r_i \times (\vec v_i-\vec u_{C_i})\right]|_z\,, \\
B&=\sum_{i=1}^{\mathcal{N}_{C_i}}\vec r_i \cdot (\vec v_i-\vec u_{C_i})\,.
\end{align}
\end{subequations}

A thermostat for translational and rotational velocity is implemented to control the temperature
of the system. Equipartition of the energy is applied to keep the same amount of energy in each degree of freedom. The temperature is defined as $k_\mathrm{B}T = \tfrac{1}{N}\sum_i \tfrac{1}{2}m_i v_i^2$ from the translational
kinetic energy, or as $k_{\mathrm{B}}T = \tfrac{1}{N}\sum_i I_i\omega_i^{2}$ from the rotational kinetic
energy in 2D, where $I_i$ is the moment of inertia of the particles and ${\omega_i}$ is the angular velocity. We employ a simple velocity rescaling thermostat that scales linear and angular velocities in a cell-wise fashion by a factor $\beta_\mathrm{th}=\sqrt{T/T_{C_i}}$ (see Eq.~\eqref{eq:LC-SRD-vel}), where $T_{C_i}$ is the instantaneous kinetic temperature in cell $C_i$.

We use two types of boundary conditions (BC), periodic BC in both the $x$ and $y$ directions for simulations of bulk systems, and a no-slip wall perpendicular to the $x$ axis (while periodic BC are implemented in the other direction) for simulations of a channel geometry. The no-slip BC are implemented with the usual bounce-back rule, that is, the velocity of the LC particle is inverted upon collision with the solid wall
\begin{equation}
\vec v_i^{\:new}=-\vec v_i^{\:old}\,.
\end{equation}
Additionally, ``ghost'' particles \cite{LamuraEPL2001} are used at the walls to fill partially occupied cells up to the average particle density.
At solid interfaces the anchoring, that is, the preferential angle between LC particles and walls needs to be specified. Homeotropic anchoring is implemented by placing ghost particles in the walls with their orientations  aligned perpendicularly to the walls.

The system is initialized by assigning the positions $\vec r_i$ and velocities $\vec v_i$. The particles are uniformly distributed in space. The linear and angular velocities are assigned so that they are distributed according to a Maxwell distribution at the target temperature, and that the equipartition theorem is obeyed.

We perform all our simulations with the following parameters: the SRD rotation angle is $\alpha = 120\degree$, particle mass $m_i = 1$, moment of inertia $I_i=1$, and the elastic and molecular relaxation constants are $\gamma_{_\mathrm{EL}} = 10^{-4}$ and $\gamma = 8\times10^{-4}$. 
It is a common choice to set the SRD grid size $\delta x=1$ and the timestep $\delta t = 1$. 
At each time step the grid is shifted by a random displacement vector with components uniformly distributed in the interval $[-\delta x/2, \delta x/2]$.
In the following we measure the temperature $T$ in units of $m_iV^2_\mathrm{max}/k_\mathrm{B}$, where $V_\mathrm{max}=\delta x/\delta t$ is the maximum propagation speed of a particle (related to the Courant condition).
The mean free path $\lambda=\delta t\sqrt{k_\mathrm{B}T/m_i}$ is typically smaller than one but the grid-shift method avoids the build-up of spurious correlations.



\section{Applications}
\label{sec:applications}

\subsection{Isotropic-Nematic Phase Transition}
\label{sec:Phase_transition}

We perform simulations of a system in equilibrium at fixed $T$ with periodic BC in both $x$ and $y$ directions. 
A useful way to characterize the degree of ordering of a nematic LC is to define an order parameter that is zero in the isotropic phase and one in the nematic phase. It is common to take the largest eigenvalue $S$ of the nematic order tensor
\begin{equation}
\mathbf{Q} = \frac{1}{N}\sum_{i=1}^{N} \left(\frac{3}{2}\hat{d_{i}}\otimes\hat{d_{i}}-\frac{1}{2}\mathbf{I}\right)\,,
\label{Order_tensor}
\end{equation}
which is a traceless, second-order tensor, and where $\otimes$ is the dyadic product and $\mathbf{I}$ is the unit tensor.

\begin{figure}
\includegraphics[width=1.0\columnwidth]{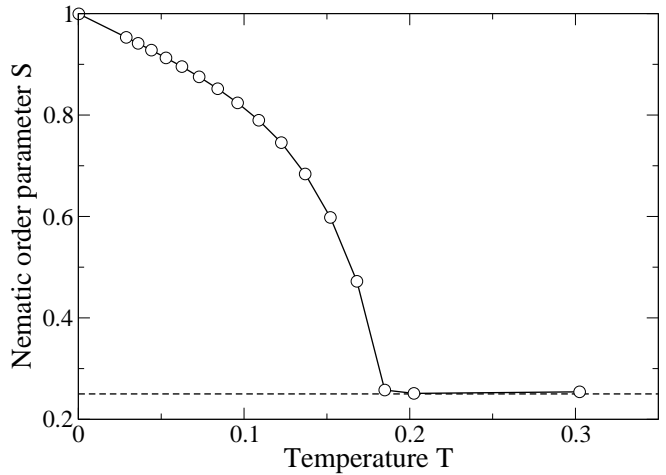}
\caption{Nematic order parameter as a function of thermostat temperature. The solid line
is just a guide for the eye. A continuous isotropic-nematic phase transition occurs at $T\simeq0.18$.}
\label{fig1}
\end{figure}

We consider a system of $N =150\,000$ particles, and of size $L_x=L_y=50$, subdivided with a $50\times50$ grid, thus the average number of particles per cell $\langle\mathcal{N}_{C_i}\rangle=60$.
 Figure~\ref{fig1} shows the dependence of the nematic order parameter $S$ on $T$, at fixed density.
At low $T$ the nematic order parameter is very close to one, indicating nearly perfect nematic alignment. As $T$ increases $S$ decreases continuously until it reaches the constant value $1/4$, which characterizes the isotropic phase in 2D~\cite{note_MaierSaupe}. We find that the isotropic-nematic transition occurs at $T\simeq0.18$.
A LC system in two dimensions or higher undergoes a temperature-driven
isotropic-nematic phase transition. This phase transition is continuous in 2D, as expected for a system with the symmetry of the XY model (though there is some controversy \cite{Vink-PRL-2007,Jordens-NatComm-2013,Marucci-Macromol-1989}).


Contrary to the hydrodynamic Ericksen-Leslie model, our SRD hybrid model can reproduce the isotropic-nematic
phase transition. The reason for that is the particle nature of our model. Different parts of the system may have different local orientations and thus, at high enough $T$, the system is globally disordered. 


\subsection{Dynamics of Topological Defects}
\label{sec:Move_defects}

During a quench from the isotropic to the nematic phase a LC substance develops many topological defects~\cite{demus1978} which influence the kinetics of the phase transition. 
Solid boundaries imposing geometric or energetic (anchoring) constraints, or external fields in general  may induce different, local  orientations in a nematic LC. Because of the conflict between these local orientations topological defects are produced. These are regions of the fluid where the local nematic order parameter is vanishing and the director field is undefined~\cite{Chaikin}.

\begin{figure}
\includegraphics[width=1.0\columnwidth]{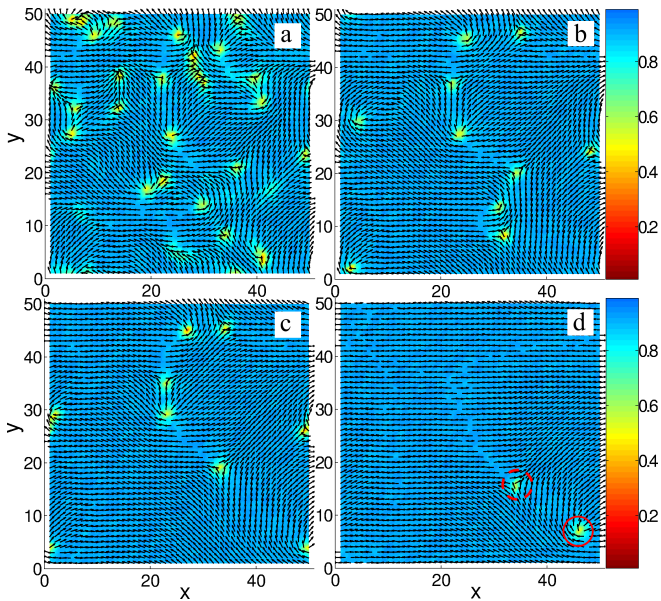}
\caption{Snapshots of the system at different times showing the temporal evolution of local nematic order parameter (color) and local director (black dashes). As the system is quenched from the isotropic to the nematic phase topological defects are formed. 
In panel (d) the red dashed circle marks the $q = -1/2$ topological defect, while the red solid circle indicates the $q = +1/2$ defect.}
\label{fig2}
\end{figure}

\begin{figure}
\centering
\includegraphics[width=1.1\columnwidth]{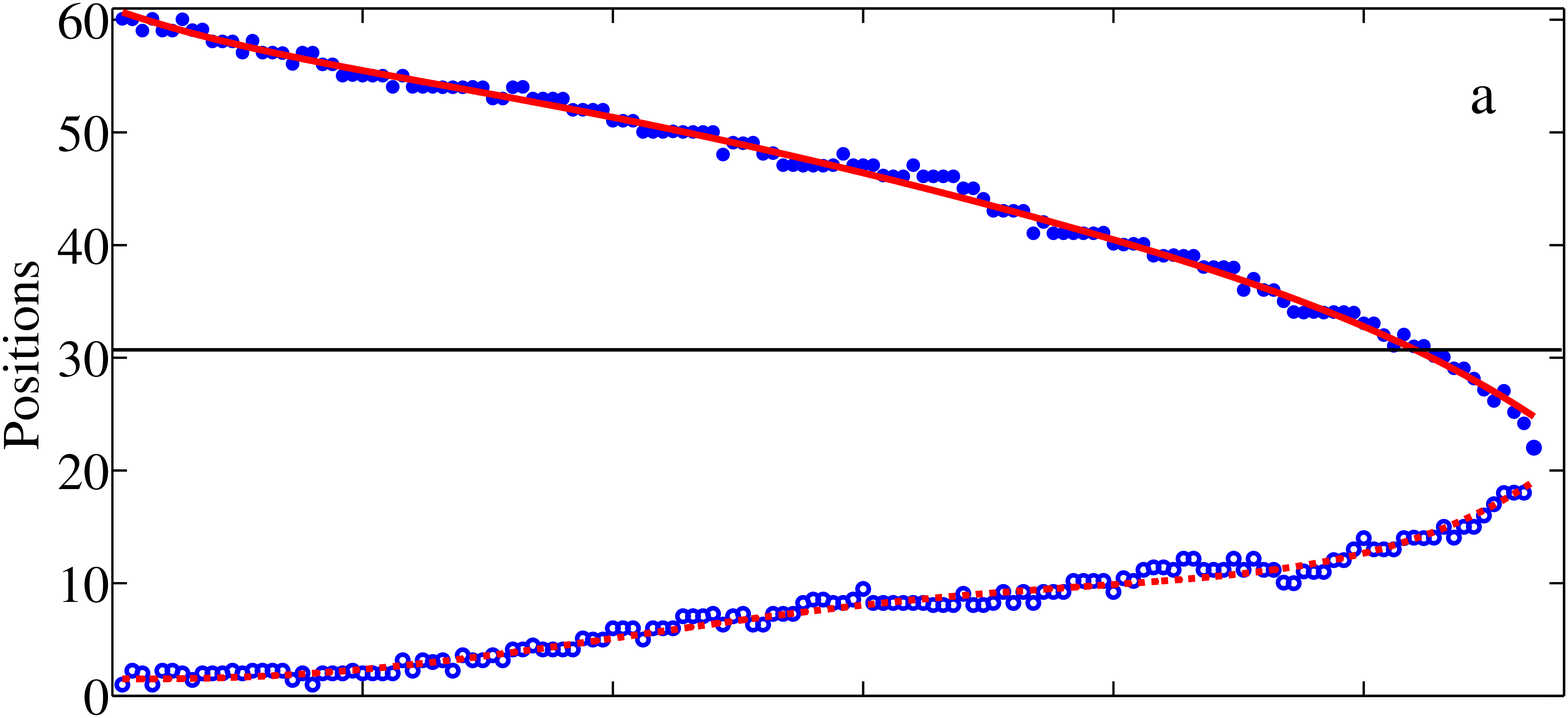}
\includegraphics[width=1.1\columnwidth]{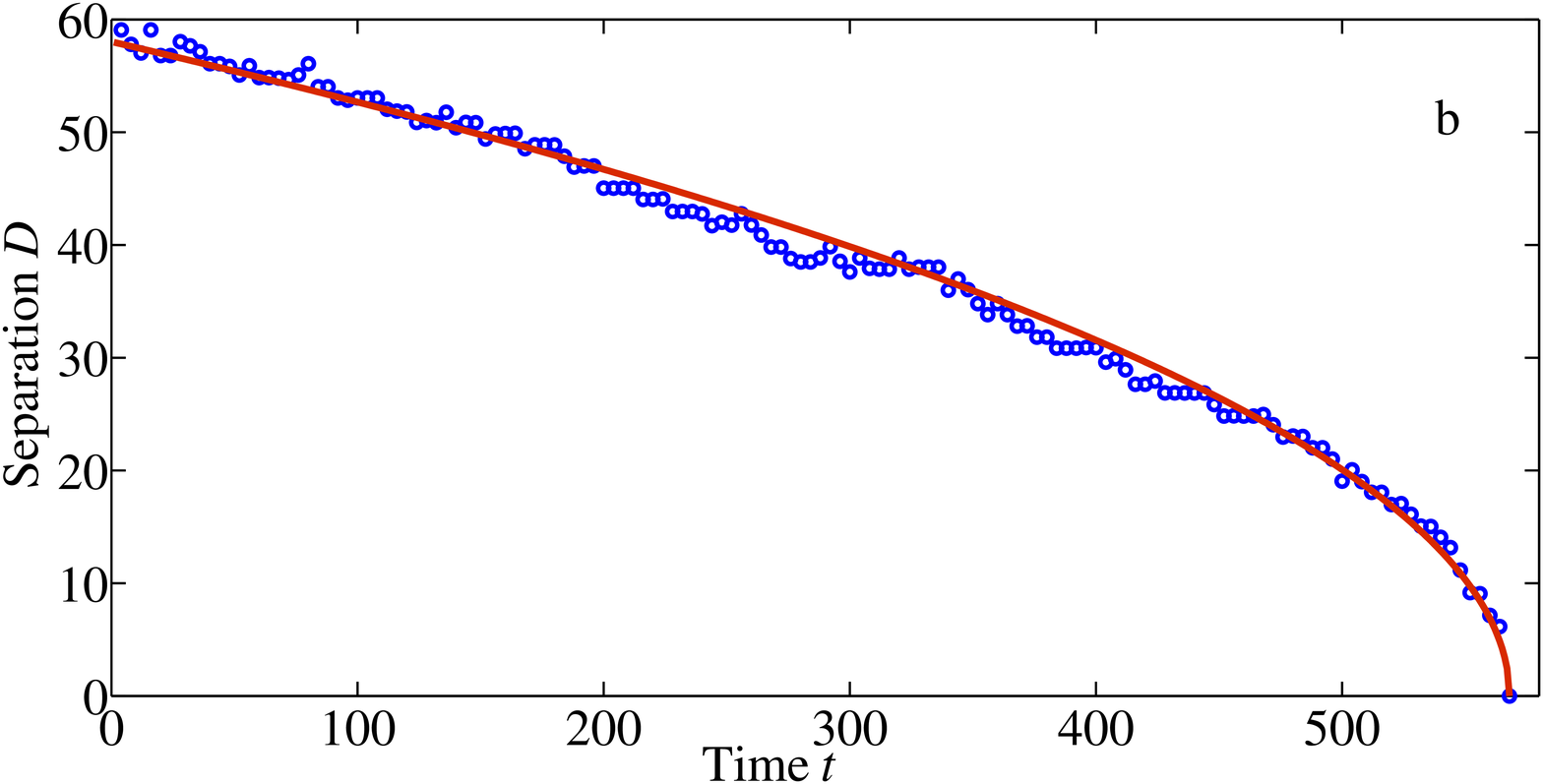}
\caption{The position of topological charges as function of time is shown in the upper panel. The blue dots
(empty circles) represent the position of positive (negative) charge. The red solid (dashed) line are the fits
of the positive (negative) charge. The lower panel shows their separation as a function of time. The blue empty
circles are the direct subtractions of positions and the red line is the fitting curve using Eq.~(\ref{distance_charges}).}
\label{fig3}
\end{figure}

Although there were attempts of using Ericksen-Leslie model to simulate topological
charges \cite{Liu:2007}, those structures  cannot be properly treated  since the nematic order parameter is
essentially constant in that approach. However in our model, the nematic order parameter in the macroscopic scale is calculated from the mesoscopic SRD particle configurations, hence the formation of topological defects is an essential feature in this particle-based approach.

We investigate the generation of topological defects and their dynamics by quenching the system from a high value of $T$, well in the isotropic phase, into the nematic phase at $T = 0.053$, with an equilibrium value of $S = 0.91$.
Figure~\ref{fig2} shows the temporal evolution of the topological defects formed during the quench. A large number of $\pm1/2$ charges are generated. The total topological charge of the system is zero, and, by conservation of charge, an equal number of positive and negative charges are present. Immediately after the initial quench the SRD particles start to align with their neighbors, due to the torque induced by the molecular field $f(\vec d)$.
After some time the particles start to form larger nematic domains. At this
stage (Fig.~\ref{fig2}(a)), there are many topological defects, carrying topological charges of $q=+1/2$
and $q=-1/2$. At a later stage, those charges with opposite signs are attracted to each others and
eventually annihilated with each others (Fig.~\ref{fig2}(b-d)), leaving the area charge-free as the initial configuration. 

The speeds of topological defects depend on the charges they carry, that is, defects with charge $q=+1/2$ 
move faster than $q=-1/2$ charges; this has been verified both by numerical~\cite{FukudaEPJB1998,Yeomans:2002,SvensekPRE2002} and experimental work~\cite{BlancPRL2005}. We find a similar behavior in our simulations. Figure~\ref{fig3}(a)
shows the time dependence of the positions of two topological charges. Initially, the charges approach each
others with rather slow velocities, but when the two charges are very close they accelerate until they annihilate. It is clear the velocity of the positive charge is higher than the negative one. The reason for this asymmetry is that the velocity of the defect core depends on the sign of the spatial derivative of the director field $\nabla \vec d$. Similar result has been reported in~\cite{Yeomans:2002} where lattice Boltzmann simulations of a tensorial formulation of nematodynamics were performed. 

To further verify the validity of our model we test the time dependence of the separation $D$ between opposite charges. 
Denniston predicted~\cite{Denniston:1996} a simple scaling law
\begin{equation}
D(t) = c(t_{a}-t)^{1/2}
\label{distance_charges}
\end{equation}
where $t_{a}$ is the annihilation time and $c$ is a constant. Figure~\ref{fig3}(b) shows that the separation between opposite charges does indeed follow the behavior predicted by Eq.~\eqref{distance_charges}. Although the scaling law~\eqref{distance_charges} was derived in conditions of no backflow~\cite{Denniston:1996}, it is still valid when the coupling of flow and director field is present, as observed in the tensorial treatments of nematodynamics~\cite{Denniston-EPL-2000}. Then we conclude that our particle-based approach to nematodynamics can correctly describe regions of the fluid with strong gradients of the nematic order parameter, such as topological defects.

\subsection{Shear flow}
\label{sec:shear_flow}

\begin{figure*}
\includegraphics[width=2\columnwidth]{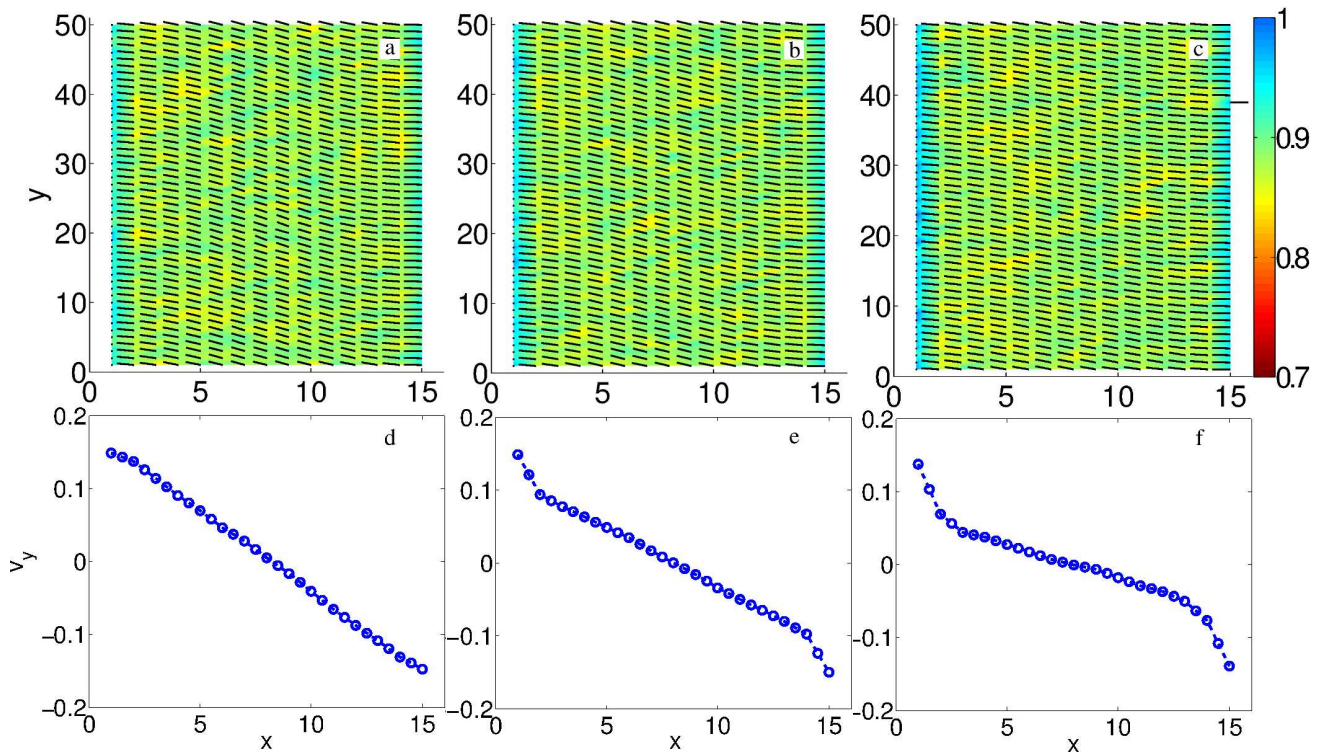}
\caption{Top row: snapshots of the nematic order parameter (color) and director fields (black dashes) under shear flow are shown. The walls at $x = 0$ and $x = L_x$ move in the $y$ direction with velocities are $v_y = 0.15$ and $v_y = -0.15$ respectively. From left to right, panels (a)-(c) show the LC for Ericksen-Leslie stress constant
$\lambda = 0$, $0.05$, $0.1$, respectively. Bottom row:  from left to right, panels (d)-(f) show the $y$-averaged flow velocity as a function of $x$, $v_{y} (x)$ for the same values of $\lambda$ as in the top row.
Shear banding caused by non-zero $\lambda$ is clearly seen by the kinks in flow profiles.}
\label{fig4}
\end{figure*}

Shear flow experiments are the canonical way to study the rheological properties of fluids. The coupling of the elastic deformations of LCs with the transport and deformation of fluid elements gives rise to a more complex situation than in the flow of an isotropic fluid. In the original formulation of Ericksen and Leslie~\cite{ericksen1959,Ericksen:1960,Ericksen:1961,leslie1966,Leslie:1968}
 the viscous response is characterized by six coefficients $\alpha_i$, $i=1...6$, called Leslie coefficients. Later, Parodi showed \cite{Parodi:1970} that from Onsager's reciprocity theorem follows $\alpha_6-\alpha_5=\alpha_2+\alpha_3$, thus the number of independent viscosity coefficients is five.

Our goal in this section is to verify if our hybrid model is capable of reproducing the known LC rheology.
To generate a shear flow in a 2D simulation domain, we consider two no-slip walls 
at $x = 0$ and $x = L_x$ and periodic BC in the $y$ direction. Because of the presence of solid walls the boundary conditions for the director field need to be specified. We employ homeotropic anchoring, that is, the LC particles prefer to orient perpendicularly to the walls. This is achieved by assigning a perpendicular orientation of the ghost particles at the walls.

The two walls move with equal and opposite speeds in the $y$ direction. After some time a stationary shear flow is generated. 
A dimensionless measure of shear is given by the Weissenberg number
$\mathcal{W}\equiv\dot{\gamma}\tau$
where $\dot{\gamma}=\partial v_y/\partial x$ is the shear rate and $\tau$ is a relaxation time. We calculate $\tau$ from the orientational correlation function $C_1(\Delta t)=\langle\vec d(t+\Delta t)\cdot\vec d(t)\rangle$, where the angle brackets indicate ensemble average. We present results for a fixed $\mathcal{W}=2.04$ at $T=0.28$ (we note that the presence of walls inducing homeotropic alignment shifts the phase diagram with respect to Fig.~\ref{fig1}, so that $T=0.28$ corresponds to the nematic state). The system of $N=75000$ particles and size $L_x=15$, $L_y=50$ is divided with a grid of $15\times 50$, thus with $\langle\mathcal{N}_{C_i}\rangle=100$.

The steady-state director fields and bulk flow profiles of sheared LC are shown in Fig.~\ref{fig4} for different Ericksen-Leslie
stress constant $\lambda$. 
For the case of $\lambda = 0$
 the flow profile is a straight line as for a Newtonian fluid.
For higher values of $\lambda$ we observe a shear banding effect. By increasing $\lambda$ the flow profile starts to have a kink near the walls.
Shear banding is a nonequilbrium transition to a state where regions with different shear rates coexist, thus visible through different slopes of the velocity profile $v_y(x)$ in Fig.~\ref{fig4}.  Nematic LCs can produce shear bands~\cite{Mather-Macromol-1997,Olmsted-Rheo-2008}. The existence of this phenomenon is a consequence of the competition between elastic energy and viscous dissipation, which in turn
is caused by the feedback from LC orientation to the flow induced by $\bm{\pi}$.

The local director field $\vec d(\vec r)$ (represented as black dashes in Fig.~\ref{fig4}(a)-(c))shows that while the director is aligned homeotropically at the walls, thus satisfying the BC, it is tilted in the central region of the channel. This tilt angle is due to a well-known flow alignment mechanism in LCs.
The local nematic order map (color code in Fig.~\ref{fig4}(a)-(c)) shows a subtle change from a high degree of nematic alignment at the walls, induced by the anchoring conditions, to a slightly lower value in the central region of the channel.

\section{Conclusions}
\label{sec:Discussions}

We have introduced a new mesoscopic LC model. Our model is based on the stochastic rotational
dynamics scheme, which is a particle-based algorithm that ignores the computationally heavy molecular interactions
but correctly generates the hydrodynamic modes described by the Navier-Stokes equations. 

The model introduced here fills an important gap for mesoscopic simulation techniques of LCs. 
Thermal fluctuations are explicitly present, which are of fundamental importance in both thermodynamic and dynamic processes, but are neglected in other popular schemes, such as the lattice Boltzmann approach. Furthermore, the mesoscopic scale and the hydrodynamic behavior is directly addressed. 

We have shown that this model can be used to study various aspects of the physics of liquid crystals. We have considered three study cases. First, we have found that our model system undergoes an equilibrium phase transition from nematic to isotropic as the temperature increases. The transition is found to be continuous, as it is expected to be in 2D. Second, we have studied the nonequilibrium dynamics of topological defects emerging in a quenched LC. Topological defects with $\pm\tfrac{1}{2}$ charge are formed but quickly annihilate with each other. The temporal dependence of the distance between two opposite charges is found to match remarkably well the theoretically predicted power law. Third, we have considered a shear flow situation and found that the LC system develops shear bands as the coupling parameter between flow and director reorientation is increased.  
We conclude that the model captures the non-Newtonian character of LC rheology.

This hybrid algorithm can be easily applied to complex geometries of the confining walls. For the sake of simplicity we restricted the present work to 2D. However, any realistic implementation requires a 3D setup. A generalization of the model to 3D is under way and will be discussed elsewhere. 

We gratefully acknowledge helpful conversations with Martin Brinkmann, Stephan Herminghaus and Thomas Hiller. 


\begin{thebibliography}{59}%
\makeatletter
\providecommand \@ifxundefined [1]{%
 \@ifx{#1\undefined}
}%
\providecommand \@ifnum [1]{%
 \ifnum #1\expandafter \@firstoftwo
 \else \expandafter \@secondoftwo
 \fi
}%
\providecommand \@ifx [1]{%
 \ifx #1\expandafter \@firstoftwo
 \else \expandafter \@secondoftwo
 \fi
}%
\providecommand \natexlab [1]{#1}%
\providecommand \enquote  [1]{``#1''}%
\providecommand \bibnamefont  [1]{#1}%
\providecommand \bibfnamefont [1]{#1}%
\providecommand \citenamefont [1]{#1}%
\providecommand \href@noop [0]{\@secondoftwo}%
\providecommand \href [0]{\begingroup \@sanitize@url \@href}%
\providecommand \@href[1]{\@@startlink{#1}\@@href}%
\providecommand \@@href[1]{\endgroup#1\@@endlink}%
\providecommand \@sanitize@url [0]{\catcode `\\12\catcode `\$12\catcode
  `\&12\catcode `\#12\catcode `\^12\catcode `\_12\catcode `\%12\relax}%
\providecommand \@@startlink[1]{}%
\providecommand \@@endlink[0]{}%
\providecommand \url  [0]{\begingroup\@sanitize@url \@url }%
\providecommand \@url [1]{\endgroup\@href {#1}{\urlprefix }}%
\providecommand \urlprefix  [0]{URL }%
\providecommand \Eprint [0]{\href }%
\providecommand \doibase [0]{http://dx.doi.org/}%
\providecommand \selectlanguage [0]{\@gobble}%
\providecommand \bibinfo  [0]{\@secondoftwo}%
\providecommand \bibfield  [0]{\@secondoftwo}%
\providecommand \translation [1]{[#1]}%
\providecommand \BibitemOpen [0]{}%
\providecommand \bibitemStop [0]{}%
\providecommand \bibitemNoStop [0]{.\EOS\space}%
\providecommand \EOS [0]{\spacefactor3000\relax}%
\providecommand \BibitemShut  [1]{\csname bibitem#1\endcsname}%
\let\auto@bib@innerbib\@empty
\bibitem [{\citenamefont {de~Gennes}\ and\ \citenamefont
  {Prost}(1993)}]{deGennes:1993}%
  \BibitemOpen
  \bibfield  {author} {\bibinfo {author} {\bibfnamefont {P.~G.}\ \bibnamefont
  {de~Gennes}}\ and\ \bibinfo {author} {\bibfnamefont {J.}~\bibnamefont
  {Prost}},\ }\href@noop {} {\emph {\bibinfo {title} {The Physics of Liquid
  Crystals}}}\ (\bibinfo  {publisher} {Clarendon Press Oxford},\ \bibinfo
  {year} {1993})\BibitemShut {NoStop}%
\bibitem [{\citenamefont {Humar}\ \emph {et~al.}(2009)\citenamefont {Humar},
  \citenamefont {Ravnik}, \citenamefont {Pajk},\ and\ \citenamefont
  {Mu{\v{s}}evi{\v{c}}}}]{humarNatPhot2009}%
  \BibitemOpen
  \bibfield  {author} {\bibinfo {author} {\bibfnamefont {M.}~\bibnamefont
  {Humar}}, \bibinfo {author} {\bibfnamefont {M.}~\bibnamefont {Ravnik}},
  \bibinfo {author} {\bibfnamefont {S.}~\bibnamefont {Pajk}}, \ and\ \bibinfo
  {author} {\bibfnamefont {I.}~\bibnamefont {Mu{\v{s}}evi{\v{c}}}},\
  }\href@noop {} {\bibfield  {journal} {\bibinfo  {journal} {Nature Photonics}\
  }\textbf {\bibinfo {volume} {3}},\ \bibinfo {pages} {595} (\bibinfo {year}
  {2009})}\BibitemShut {NoStop}%
\bibitem [{\citenamefont {Humar}\ and\ \citenamefont
  {Mu{\v{s}}evi{\v{c}}}(2010)}]{humar-OE-2010}%
  \BibitemOpen
  \bibfield  {author} {\bibinfo {author} {\bibfnamefont {M.}~\bibnamefont
  {Humar}}\ and\ \bibinfo {author} {\bibfnamefont {I.}~\bibnamefont
  {Mu{\v{s}}evi{\v{c}}}},\ }\href@noop {} {\bibfield  {journal} {\bibinfo
  {journal} {Opt. Express}\ }\textbf {\bibinfo {volume} {18}},\ \bibinfo
  {pages} {26995} (\bibinfo {year} {2010})}\BibitemShut {NoStop}%
\bibitem [{\citenamefont {Peddireddy}\ \emph {et~al.}(2013)\citenamefont
  {Peddireddy}, \citenamefont {Jampani}, \citenamefont {Thutupalli},
  \citenamefont {Herminghaus}, \citenamefont {Bahr},\ and\ \citenamefont
  {Mu{\v{s}}evi{\v{c}}}}]{peddireddy-OE-2013}%
  \BibitemOpen
  \bibfield  {author} {\bibinfo {author} {\bibfnamefont {K.}~\bibnamefont
  {Peddireddy}}, \bibinfo {author} {\bibfnamefont {V.~S.~R.}\ \bibnamefont
  {Jampani}}, \bibinfo {author} {\bibfnamefont {S.}~\bibnamefont {Thutupalli}},
  \bibinfo {author} {\bibfnamefont {S.}~\bibnamefont {Herminghaus}}, \bibinfo
  {author} {\bibfnamefont {C.}~\bibnamefont {Bahr}}, \ and\ \bibinfo {author}
  {\bibfnamefont {I.}~\bibnamefont {Mu{\v{s}}evi{\v{c}}}},\ }\href@noop {}
  {\bibfield  {journal} {\bibinfo  {journal} {Opt. Express}\ }\textbf {\bibinfo
  {volume} {21}},\ \bibinfo {pages} {30233} (\bibinfo {year}
  {2013})}\BibitemShut {NoStop}%
\bibitem [{\citenamefont {Amann}\ \emph {et~al.}(2013)\citenamefont {Amann},
  \citenamefont {Dold},\ and\ \citenamefont {Kailer}}]{Amann-2013}%
  \BibitemOpen
  \bibfield  {author} {\bibinfo {author} {\bibfnamefont {T.}~\bibnamefont
  {Amann}}, \bibinfo {author} {\bibfnamefont {C.}~\bibnamefont {Dold}}, \ and\
  \bibinfo {author} {\bibfnamefont {A.}~\bibnamefont {Kailer}},\ }\href@noop {}
  {\bibfield  {journal} {\bibinfo  {journal} {Tribol. Int.}\ }\textbf {\bibinfo
  {volume} {65}},\ \bibinfo {pages} {3} (\bibinfo {year} {2013})}\BibitemShut
  {NoStop}%
\bibitem [{\citenamefont {Stewart}(2003)}]{stewart2003liquid}%
  \BibitemOpen
  \bibfield  {author} {\bibinfo {author} {\bibfnamefont {G.~T.}\ \bibnamefont
  {Stewart}},\ }\href@noop {} {\bibfield  {journal} {\bibinfo  {journal} {Liq.
  Cryst.}\ }\textbf {\bibinfo {volume} {30}},\ \bibinfo {pages} {541} (\bibinfo
  {year} {2003})}\BibitemShut {NoStop}%
\bibitem [{\citenamefont {Stewart}(2004)}]{stewart2004liquid}%
  \BibitemOpen
  \bibfield  {author} {\bibinfo {author} {\bibfnamefont {G.~T.}\ \bibnamefont
  {Stewart}},\ }\href@noop {} {\bibfield  {journal} {\bibinfo  {journal} {Liq.
  Cryst.}\ }\textbf {\bibinfo {volume} {31}},\ \bibinfo {pages} {443} (\bibinfo
  {year} {2004})}\BibitemShut {NoStop}%
\bibitem [{\citenamefont {Stephen}\ and\ \citenamefont
  {Straley}(1974)}]{Stephen-1974}%
  \BibitemOpen
  \bibfield  {author} {\bibinfo {author} {\bibfnamefont {M.~J.}\ \bibnamefont
  {Stephen}}\ and\ \bibinfo {author} {\bibfnamefont {J.~P.}\ \bibnamefont
  {Straley}},\ }\href {\doibase 10.1103/RevModPhys.46.617} {\bibfield
  {journal} {\bibinfo  {journal} {Rev. Mod. Phys.}\ }\textbf {\bibinfo {volume}
  {46}},\ \bibinfo {pages} {617} (\bibinfo {year} {1974})}\BibitemShut
  {NoStop}%
\bibitem [{\citenamefont {Sengupta}\ \emph
  {et~al.}(2013{\natexlab{a}})\citenamefont {Sengupta}, \citenamefont {Tkalec},
  \citenamefont {Ravnik}, \citenamefont {Yeomans}, \citenamefont {Bahr},\ and\
  \citenamefont {Herminghaus}}]{Sengupta:2013}%
  \BibitemOpen
  \bibfield  {author} {\bibinfo {author} {\bibfnamefont {A.}~\bibnamefont
  {Sengupta}}, \bibinfo {author} {\bibfnamefont {U.}~\bibnamefont {Tkalec}},
  \bibinfo {author} {\bibfnamefont {M.}~\bibnamefont {Ravnik}}, \bibinfo
  {author} {\bibfnamefont {J.~M.}\ \bibnamefont {Yeomans}}, \bibinfo {author}
  {\bibfnamefont {C.}~\bibnamefont {Bahr}}, \ and\ \bibinfo {author}
  {\bibfnamefont {S.}~\bibnamefont {Herminghaus}},\ }\href@noop {} {\bibfield
  {journal} {\bibinfo  {journal} {Phys. Rev. Lett.}\ }\textbf {\bibinfo
  {volume} {110}},\ \bibinfo {pages} {048303} (\bibinfo {year}
  {2013}{\natexlab{a}})}\BibitemShut {NoStop}%
\bibitem [{\citenamefont {Sengupta}\ \emph
  {et~al.}(2013{\natexlab{b}})\citenamefont {Sengupta}, \citenamefont {Pieper},
  \citenamefont {Enderlein}, \citenamefont {Bahr},\ and\ \citenamefont
  {Herminghaus}}]{Sengupta-cyl-SM-13}%
  \BibitemOpen
  \bibfield  {author} {\bibinfo {author} {\bibfnamefont {A.}~\bibnamefont
  {Sengupta}}, \bibinfo {author} {\bibfnamefont {C.}~\bibnamefont {Pieper}},
  \bibinfo {author} {\bibfnamefont {J.}~\bibnamefont {Enderlein}}, \bibinfo
  {author} {\bibfnamefont {C.}~\bibnamefont {Bahr}}, \ and\ \bibinfo {author}
  {\bibfnamefont {S.}~\bibnamefont {Herminghaus}},\ }\href {\doibase
  10.1039/C2SM27337C} {\bibfield  {journal} {\bibinfo  {journal} {Soft Matter}\
  }\textbf {\bibinfo {volume} {9}},\ \bibinfo {pages} {1937} (\bibinfo {year}
  {2013}{\natexlab{b}})}\BibitemShut {NoStop}%
\bibitem [{\citenamefont {Sengupta}\ \emph
  {et~al.}(2013{\natexlab{c}})\citenamefont {Sengupta}, \citenamefont {Bahr},\
  and\ \citenamefont {Herminghaus}}]{Sengupta-SM-13}%
  \BibitemOpen
  \bibfield  {author} {\bibinfo {author} {\bibfnamefont {A.}~\bibnamefont
  {Sengupta}}, \bibinfo {author} {\bibfnamefont {C.}~\bibnamefont {Bahr}}, \
  and\ \bibinfo {author} {\bibfnamefont {S.}~\bibnamefont {Herminghaus}},\
  }\href {\doibase 10.1039/C3SM50677K} {\bibfield  {journal} {\bibinfo
  {journal} {Soft Matter}\ }\textbf {\bibinfo {volume} {9}},\ \bibinfo {pages}
  {7251} (\bibinfo {year} {2013}{\natexlab{c}})}\BibitemShut {NoStop}%
\bibitem [{\citenamefont {Sengupta}(2014)}]{Sengupta-LC-14}%
  \BibitemOpen
  \bibfield  {author} {\bibinfo {author} {\bibfnamefont {A.}~\bibnamefont
  {Sengupta}},\ }\href {\doibase 10.1080/02678292.2013.807939} {\bibfield
  {journal} {\bibinfo  {journal} {Liquid Crystals}\ }\textbf {\bibinfo {volume}
  {41}},\ \bibinfo {pages} {290} (\bibinfo {year} {2014})}\BibitemShut
  {NoStop}%
\bibitem [{\citenamefont {Sengupta}(2013)}]{Sengupta-IJMS-13}%
  \BibitemOpen
  \bibfield  {author} {\bibinfo {author} {\bibfnamefont {A.}~\bibnamefont
  {Sengupta}},\ }\href@noop {} {\bibfield  {journal} {\bibinfo  {journal} {Int.
  J. Mol. Sci.}\ }\textbf {\bibinfo {volume} {14}},\ \bibinfo {pages} {22826}
  (\bibinfo {year} {2013})}\BibitemShut {NoStop}%
\bibitem [{\citenamefont {Parodi}(1970)}]{Parodi:1970}%
  \BibitemOpen
  \bibfield  {author} {\bibinfo {author} {\bibfnamefont {O.}~\bibnamefont
  {Parodi}},\ }\href@noop {} {\bibfield  {journal} {\bibinfo  {journal}
  {Journal de Physique}\ }\textbf {\bibinfo {volume} {31}},\ \bibinfo {pages}
  {581} (\bibinfo {year} {1970})}\BibitemShut {NoStop}%
\bibitem [{\citenamefont {Oseen}(1933)}]{Oseen}%
  \BibitemOpen
  \bibfield  {author} {\bibinfo {author} {\bibfnamefont {C.~W.}\ \bibnamefont
  {Oseen}},\ }\href {\doibase 10.1039/TF9332900883} {\bibfield  {journal}
  {\bibinfo  {journal} {Trans. Faraday Soc.}\ }\textbf {\bibinfo {volume}
  {29}},\ \bibinfo {pages} {883} (\bibinfo {year} {1933})}\BibitemShut
  {NoStop}%
\bibitem [{\citenamefont {Zocher}(1933)}]{Zocher}%
  \BibitemOpen
  \bibfield  {author} {\bibinfo {author} {\bibfnamefont {H.}~\bibnamefont
  {Zocher}},\ }\href {\doibase 10.1039/TF9332900945} {\bibfield  {journal}
  {\bibinfo  {journal} {Trans. Faraday Soc.}\ }\textbf {\bibinfo {volume}
  {29}},\ \bibinfo {pages} {945} (\bibinfo {year} {1933})}\BibitemShut
  {NoStop}%
\bibitem [{\citenamefont {Frank}(1958)}]{Frank:1958}%
  \BibitemOpen
  \bibfield  {author} {\bibinfo {author} {\bibfnamefont {F.~C.}\ \bibnamefont
  {Frank}},\ }\href@noop {} {\bibfield  {journal} {\bibinfo  {journal}
  {Discuss. Faraday Soc.}\ }\textbf {\bibinfo {volume} {25}},\ \bibinfo {pages}
  {19} (\bibinfo {year} {1958})}\BibitemShut {NoStop}%
\bibitem [{\citenamefont {Anzelius}(1931)}]{Anzelius}%
  \BibitemOpen
  \bibfield  {author} {\bibinfo {author} {\bibfnamefont {A.}~\bibnamefont
  {Anzelius}},\ }\href@noop {} {\bibfield  {journal} {\bibinfo  {journal}
  {Uppsala Univ. Arrskr., Mat. och Naturve}\ ,\ \bibinfo {pages} {1}} (\bibinfo
  {year} {1931})}\BibitemShut {NoStop}%
\bibitem [{\citenamefont {Ericksen}(1959)}]{ericksen1959}%
  \BibitemOpen
  \bibfield  {author} {\bibinfo {author} {\bibfnamefont {J.}~\bibnamefont
  {Ericksen}},\ }\href@noop {} {\bibfield  {journal} {\bibinfo  {journal}
  {Arch. Ration. Mech. An.}\ }\textbf {\bibinfo {volume} {4}},\ \bibinfo
  {pages} {231} (\bibinfo {year} {1959})}\BibitemShut {NoStop}%
\bibitem [{\citenamefont {Ericksen}(1960)}]{Ericksen:1960}%
  \BibitemOpen
  \bibfield  {author} {\bibinfo {author} {\bibfnamefont {J.~L.}\ \bibnamefont
  {Ericksen}},\ }\href {\doibase http://dx.doi.org/10.1122/1.548864} {\bibfield
   {journal} {\bibinfo  {journal} {Trans. Soc. Rheol.}\ }\textbf {\bibinfo
  {volume} {4}},\ \bibinfo {pages} {29} (\bibinfo {year} {1960})}\BibitemShut
  {NoStop}%
\bibitem [{\citenamefont {Ericksen}(1961)}]{Ericksen:1961}%
  \BibitemOpen
  \bibfield  {author} {\bibinfo {author} {\bibfnamefont {J.~L.}\ \bibnamefont
  {Ericksen}},\ }\href@noop {} {\bibfield  {journal} {\bibinfo  {journal}
  {Trans. Soc. Rheol.}\ }\textbf {\bibinfo {volume} {5}},\ \bibinfo {pages}
  {23} (\bibinfo {year} {1961})}\BibitemShut {NoStop}%
\bibitem [{\citenamefont {Leslie}(1966)}]{leslie1966}%
  \BibitemOpen
  \bibfield  {author} {\bibinfo {author} {\bibfnamefont {F.~M.}\ \bibnamefont
  {Leslie}},\ }\href@noop {} {\bibfield  {journal} {\bibinfo  {journal} {Q. J.
  Mech. Appl. Math.}\ }\textbf {\bibinfo {volume} {19}},\ \bibinfo {pages}
  {357} (\bibinfo {year} {1966})}\BibitemShut {NoStop}%
\bibitem [{\citenamefont {Leslie}(1968)}]{Leslie:1968}%
  \BibitemOpen
  \bibfield  {author} {\bibinfo {author} {\bibfnamefont {F.~M.}\ \bibnamefont
  {Leslie}},\ }\href@noop {} {\bibfield  {journal} {\bibinfo  {journal} {Arch.
  Rat. Mech. An.}\ }\textbf {\bibinfo {volume} {28}},\ \bibinfo {pages} {265}
  (\bibinfo {year} {1968})}\BibitemShut {NoStop}%
\bibitem [{\citenamefont {Beris}\ and\ \citenamefont
  {Edwards}(1994)}]{beris1994thermodynamics}%
  \BibitemOpen
  \bibfield  {author} {\bibinfo {author} {\bibfnamefont {A.}~\bibnamefont
  {Beris}}\ and\ \bibinfo {author} {\bibfnamefont {S.}~\bibnamefont
  {Edwards}},\ }\href@noop {} {\emph {\bibinfo {title} {Thermodynamics of
  flowing fluids}}}\ (\bibinfo  {publisher} {Oxford University Press, New
  York},\ \bibinfo {year} {1994})\BibitemShut {NoStop}%
\bibitem [{\citenamefont {Qian}\ and\ \citenamefont
  {Sheng}(1998)}]{Qian-Sheng-PRE}%
  \BibitemOpen
  \bibfield  {author} {\bibinfo {author} {\bibfnamefont {T.}~\bibnamefont
  {Qian}}\ and\ \bibinfo {author} {\bibfnamefont {P.}~\bibnamefont {Sheng}},\
  }\href {\doibase 10.1103/PhysRevE.58.7475} {\bibfield  {journal} {\bibinfo
  {journal} {Phys. Rev. E}\ }\textbf {\bibinfo {volume} {58}},\ \bibinfo
  {pages} {7475} (\bibinfo {year} {1998})}\BibitemShut {NoStop}%
\bibitem [{\citenamefont {McNamara}\ and\ \citenamefont
  {Zanetti}(1988)}]{McnamaraZanettiPRL1988}%
  \BibitemOpen
  \bibfield  {author} {\bibinfo {author} {\bibfnamefont {G.~R.}\ \bibnamefont
  {McNamara}}\ and\ \bibinfo {author} {\bibfnamefont {G.}~\bibnamefont
  {Zanetti}},\ }\href {\doibase 10.1103/PhysRevLett.61.2332} {\bibfield
  {journal} {\bibinfo  {journal} {Phys. Rev. Lett.}\ }\textbf {\bibinfo
  {volume} {61}},\ \bibinfo {pages} {2332} (\bibinfo {year}
  {1988})}\BibitemShut {NoStop}%
\bibitem [{\citenamefont {Denniston}\ \emph {et~al.}(2000)\citenamefont
  {Denniston}, \citenamefont {Orlandini},\ and\ \citenamefont
  {Yeomans}}]{Denniston-EPL-2000}%
  \BibitemOpen
  \bibfield  {author} {\bibinfo {author} {\bibfnamefont {C.}~\bibnamefont
  {Denniston}}, \bibinfo {author} {\bibfnamefont {E.}~\bibnamefont
  {Orlandini}}, \ and\ \bibinfo {author} {\bibfnamefont {J.~M.}\ \bibnamefont
  {Yeomans}},\ }\href@noop {} {\bibfield  {journal} {\bibinfo  {journal}
  {Europhys. Lett.}\ }\textbf {\bibinfo {volume} {52}},\ \bibinfo {pages} {481}
  (\bibinfo {year} {2000})}\BibitemShut {NoStop}%
\bibitem [{\citenamefont {Care}\ \emph {et~al.}(2000)\citenamefont {Care},
  \citenamefont {Halliday},\ and\ \citenamefont {Good}}]{CareJPCM2000}%
  \BibitemOpen
  \bibfield  {author} {\bibinfo {author} {\bibfnamefont {C.~M.}\ \bibnamefont
  {Care}}, \bibinfo {author} {\bibfnamefont {I.}~\bibnamefont {Halliday}}, \
  and\ \bibinfo {author} {\bibfnamefont {K.}~\bibnamefont {Good}},\ }\href
  {http://stacks.iop.org/0953-8984/12/i=43/a=101} {\bibfield  {journal}
  {\bibinfo  {journal} {J. Phys.: Condens. Matter}\ }\textbf {\bibinfo {volume}
  {12}},\ \bibinfo {pages} {L665} (\bibinfo {year} {2000})}\BibitemShut
  {NoStop}%
\bibitem [{\citenamefont {Care}\ \emph {et~al.}(2003)\citenamefont {Care},
  \citenamefont {Halliday}, \citenamefont {Good},\ and\ \citenamefont
  {Lishchuk}}]{CarePRE2003}%
  \BibitemOpen
  \bibfield  {author} {\bibinfo {author} {\bibfnamefont {C.~M.}\ \bibnamefont
  {Care}}, \bibinfo {author} {\bibfnamefont {I.}~\bibnamefont {Halliday}},
  \bibinfo {author} {\bibfnamefont {K.}~\bibnamefont {Good}}, \ and\ \bibinfo
  {author} {\bibfnamefont {S.~V.}\ \bibnamefont {Lishchuk}},\ }\href {\doibase
  10.1103/PhysRevE.67.061703} {\bibfield  {journal} {\bibinfo  {journal} {Phys.
  Rev. E}\ }\textbf {\bibinfo {volume} {67}},\ \bibinfo {pages} {061703}
  (\bibinfo {year} {2003})}\BibitemShut {NoStop}%
\bibitem [{\citenamefont {AlSunaidi}\ \emph {et~al.}(2004)\citenamefont
  {AlSunaidi}, \citenamefont {den Otter},\ and\ \citenamefont
  {Clarke}}]{AlSunaidi2004}%
  \BibitemOpen
  \bibfield  {author} {\bibinfo {author} {\bibfnamefont {A.}~\bibnamefont
  {AlSunaidi}}, \bibinfo {author} {\bibfnamefont {W.~K.}\ \bibnamefont {den
  Otter}}, \ and\ \bibinfo {author} {\bibfnamefont {J.~H.~R.}\ \bibnamefont
  {Clarke}},\ }\href {\doibase 10.1098/rsta.2004.1414} {\bibfield  {journal}
  {\bibinfo  {journal} {Philos. Trans. R. Soc. London, Ser. A}\ }\textbf
  {\bibinfo {volume} {362}},\ \bibinfo {pages} {1773} (\bibinfo {year}
  {2004})}\BibitemShut {NoStop}%
\bibitem [{\citenamefont {Levine}\ \emph {et~al.}(2005)\citenamefont {Levine},
  \citenamefont {Gomes}, \citenamefont {Martins},\ and\ \citenamefont
  {Polimeno}}]{LevineJCP2005}%
  \BibitemOpen
  \bibfield  {author} {\bibinfo {author} {\bibfnamefont {Y.~K.}\ \bibnamefont
  {Levine}}, \bibinfo {author} {\bibfnamefont {A.~E.}\ \bibnamefont {Gomes}},
  \bibinfo {author} {\bibfnamefont {A.~F.}\ \bibnamefont {Martins}}, \ and\
  \bibinfo {author} {\bibfnamefont {A.}~\bibnamefont {Polimeno}},\ }\href
  {\doibase http://dx.doi.org/10.1063/1.1879852} {\bibfield  {journal}
  {\bibinfo  {journal} {J. Chem. Phys.}\ }\textbf {\bibinfo {volume} {122}},\
  \bibinfo {eid} {144902} (\bibinfo {year} {2005})}\BibitemShut {NoStop}%
\bibitem [{\citenamefont {Hoogerbrugge}\ and\ \citenamefont
  {Koelman}(1992)}]{HoogerbruggeEPL1992}%
  \BibitemOpen
  \bibfield  {author} {\bibinfo {author} {\bibfnamefont {P.~J.}\ \bibnamefont
  {Hoogerbrugge}}\ and\ \bibinfo {author} {\bibfnamefont {J.~M. V.~A.}\
  \bibnamefont {Koelman}},\ }\href
  {http://stacks.iop.org/0295-5075/19/i=3/a=001} {\bibfield  {journal}
  {\bibinfo  {journal} {Europhys. Lett.}\ }\textbf {\bibinfo {volume} {19}},\
  \bibinfo {pages} {155} (\bibinfo {year} {1992})}\BibitemShut {NoStop}%
\bibitem [{\citenamefont {Malevanets}\ and\ \citenamefont
  {Kapral}(1999)}]{Malevanets:1999}%
  \BibitemOpen
  \bibfield  {author} {\bibinfo {author} {\bibfnamefont {A.}~\bibnamefont
  {Malevanets}}\ and\ \bibinfo {author} {\bibfnamefont {R.}~\bibnamefont
  {Kapral}},\ }\href@noop {} {\bibfield  {journal} {\bibinfo  {journal} {J.
  Chem. Phys.}\ }\textbf {\bibinfo {volume} {110}},\ \bibinfo {pages} {8605}
  (\bibinfo {year} {1999})}\BibitemShut {NoStop}%
\bibitem [{\citenamefont {Padding}\ and\ \citenamefont
  {Louis}(2004)}]{PAddingPRL2004}%
  \BibitemOpen
  \bibfield  {author} {\bibinfo {author} {\bibfnamefont {J.~T.}\ \bibnamefont
  {Padding}}\ and\ \bibinfo {author} {\bibfnamefont {A.~A.}\ \bibnamefont
  {Louis}},\ }\href {\doibase 10.1103/PhysRevLett.93.220601} {\bibfield
  {journal} {\bibinfo  {journal} {Phys. Rev. Lett.}\ }\textbf {\bibinfo
  {volume} {93}},\ \bibinfo {pages} {220601} (\bibinfo {year}
  {2004})}\BibitemShut {NoStop}%
\bibitem [{\citenamefont {Malevanets}\ and\ \citenamefont
  {Yeomans}(2000)}]{malevanets2000dynamics}%
  \BibitemOpen
  \bibfield  {author} {\bibinfo {author} {\bibfnamefont {A.}~\bibnamefont
  {Malevanets}}\ and\ \bibinfo {author} {\bibfnamefont {J.}~\bibnamefont
  {Yeomans}},\ }\href@noop {} {\bibfield  {journal} {\bibinfo  {journal}
  {Europhys. Lett.}\ }\textbf {\bibinfo {volume} {52}},\ \bibinfo {pages} {231}
  (\bibinfo {year} {2000})}\BibitemShut {NoStop}%
\bibitem [{\citenamefont {Winkler}\ \emph {et~al.}(2004)\citenamefont
  {Winkler}, \citenamefont {Mussawisade}, \citenamefont {Ripoll},\ and\
  \citenamefont {Gompper}}]{WinklerJPCM2004}%
  \BibitemOpen
  \bibfield  {author} {\bibinfo {author} {\bibfnamefont {R.~G.}\ \bibnamefont
  {Winkler}}, \bibinfo {author} {\bibfnamefont {K.}~\bibnamefont
  {Mussawisade}}, \bibinfo {author} {\bibfnamefont {M.}~\bibnamefont {Ripoll}},
  \ and\ \bibinfo {author} {\bibfnamefont {G.}~\bibnamefont {Gompper}},\ }\href
  {http://stacks.iop.org/0953-8984/16/i=38/a=012} {\bibfield  {journal}
  {\bibinfo  {journal} {J. Phys.: Condens. Matter}\ }\textbf {\bibinfo {volume}
  {16}},\ \bibinfo {pages} {S3941} (\bibinfo {year} {2004})}\BibitemShut
  {NoStop}%
\bibitem [{\citenamefont {Noguchi}\ and\ \citenamefont
  {Gompper}(2006)}]{NoguchiJCP2006}%
  \BibitemOpen
  \bibfield  {author} {\bibinfo {author} {\bibfnamefont {H.}~\bibnamefont
  {Noguchi}}\ and\ \bibinfo {author} {\bibfnamefont {G.}~\bibnamefont
  {Gompper}},\ }\href {\doibase http://dx.doi.org/10.1063/1.2358983} {\bibfield
   {journal} {\bibinfo  {journal} {J. Chem. Phys.}\ }\textbf {\bibinfo {volume}
  {125}},\ \bibinfo {eid} {164908} (\bibinfo {year} {2006})}\BibitemShut
  {NoStop}%
\bibitem [{\citenamefont {Noguchi}\ and\ \citenamefont
  {Gompper}(2005)}]{NoguchiPNAS2005}%
  \BibitemOpen
  \bibfield  {author} {\bibinfo {author} {\bibfnamefont {H.}~\bibnamefont
  {Noguchi}}\ and\ \bibinfo {author} {\bibfnamefont {G.}~\bibnamefont
  {Gompper}},\ }\href {\doibase 10.1073/pnas.0504243102} {\bibfield  {journal}
  {\bibinfo  {journal} {Proc. Nat. Acad. Sci. USA}\ }\textbf {\bibinfo {volume}
  {102}},\ \bibinfo {pages} {14159} (\bibinfo {year} {2005})}\BibitemShut
  {NoStop}%
\bibitem [{\citenamefont {Ihle}\ and\ \citenamefont
  {Kroll}(2001)}]{IhlePRE2001}%
  \BibitemOpen
  \bibfield  {author} {\bibinfo {author} {\bibfnamefont {T.}~\bibnamefont
  {Ihle}}\ and\ \bibinfo {author} {\bibfnamefont {D.~M.}\ \bibnamefont
  {Kroll}},\ }\href {\doibase 10.1103/PhysRevE.63.020201} {\bibfield  {journal}
  {\bibinfo  {journal} {Phys. Rev. E}\ }\textbf {\bibinfo {volume} {63}},\
  \bibinfo {pages} {020201} (\bibinfo {year} {2001})}\BibitemShut {NoStop}%
\bibitem [{\citenamefont {Lebwohl}\ and\ \citenamefont
  {Lasher}(1972)}]{Lebwohl-PRA-1972}%
  \BibitemOpen
  \bibfield  {author} {\bibinfo {author} {\bibfnamefont {P.}~\bibnamefont
  {Lebwohl}}\ and\ \bibinfo {author} {\bibfnamefont {G.}~\bibnamefont
  {Lasher}},\ }\href {\doibase 10.1103/PhysRevA.6.426} {\bibfield  {journal}
  {\bibinfo  {journal} {Phys. Rev. A}\ }\textbf {\bibinfo {volume} {6}},\
  \bibinfo {pages} {426} (\bibinfo {year} {1972})}\BibitemShut {NoStop}%
\bibitem [{\citenamefont {Lin}(1989)}]{Lin-1989}%
  \BibitemOpen
  \bibfield  {author} {\bibinfo {author} {\bibfnamefont {F.-H.}\ \bibnamefont
  {Lin}},\ }\href {\doibase 10.1002/cpa.3160420605} {\bibfield  {journal}
  {\bibinfo  {journal} {Comm. Pure Appl. Math.}\ }\textbf {\bibinfo {volume}
  {42}},\ \bibinfo {pages} {789} (\bibinfo {year} {1989})}\BibitemShut
  {NoStop}%
\bibitem [{\citenamefont {Lin}\ and\ \citenamefont {Liu}(1995)}]{Lin-1995}%
  \BibitemOpen
  \bibfield  {author} {\bibinfo {author} {\bibfnamefont {F.-H.}\ \bibnamefont
  {Lin}}\ and\ \bibinfo {author} {\bibfnamefont {C.}~\bibnamefont {Liu}},\
  }\href {\doibase 10.1002/cpa.3160480503} {\bibfield  {journal} {\bibinfo
  {journal} {Comm. Pure Appl. Math.}\ }\textbf {\bibinfo {volume} {48}},\
  \bibinfo {pages} {501} (\bibinfo {year} {1995})}\BibitemShut {NoStop}%
\bibitem [{\citenamefont {Ryder}(2005)}]{Ryder2005}%
  \BibitemOpen
  \bibfield  {author} {\bibinfo {author} {\bibfnamefont {J.}~\bibnamefont
  {Ryder}},\ }\emph {\bibinfo {title} {Mesoscopic Simulations of Complex
  Fluids}},\ \href@noop {} {Ph.D. thesis},\ \bibinfo  {school} {University of
  Oxford} (\bibinfo {year} {2005})\BibitemShut {NoStop}%
\bibitem [{\citenamefont {Gompper}\ \emph {et~al.}(2009)\citenamefont
  {Gompper}, \citenamefont {Ihle}, \citenamefont {Kroll},\ and\ \citenamefont
  {Winkler}}]{Gompper:2008}%
  \BibitemOpen
  \bibfield  {author} {\bibinfo {author} {\bibfnamefont {G.}~\bibnamefont
  {Gompper}}, \bibinfo {author} {\bibfnamefont {T.}~\bibnamefont {Ihle}},
  \bibinfo {author} {\bibfnamefont {D.}~\bibnamefont {Kroll}}, \ and\ \bibinfo
  {author} {\bibfnamefont {R.}~\bibnamefont {Winkler}},\ }\href@noop {}
  {\bibfield  {journal} {\bibinfo  {journal} {Adv. Polym. Sci.}\ }\textbf
  {\bibinfo {volume} {221}},\ \bibinfo {pages} {1} (\bibinfo {year}
  {2009})}\BibitemShut {NoStop}%
\bibitem [{\citenamefont {Lamura}\ \emph {et~al.}(2001)\citenamefont {Lamura},
  \citenamefont {Gompper}, \citenamefont {Ihle},\ and\ \citenamefont
  {Kroll}}]{LamuraEPL2001}%
  \BibitemOpen
  \bibfield  {author} {\bibinfo {author} {\bibfnamefont {A.}~\bibnamefont
  {Lamura}}, \bibinfo {author} {\bibfnamefont {G.}~\bibnamefont {Gompper}},
  \bibinfo {author} {\bibfnamefont {T.}~\bibnamefont {Ihle}}, \ and\ \bibinfo
  {author} {\bibfnamefont {D.~M.}\ \bibnamefont {Kroll}},\ }\href
  {http://stacks.iop.org/0295-5075/56/i=3/a=319} {\bibfield  {journal}
  {\bibinfo  {journal} {Europhys. Lett.}\ }\textbf {\bibinfo {volume} {56}},\
  \bibinfo {pages} {319} (\bibinfo {year} {2001})}\BibitemShut {NoStop}%
\bibitem [{not()}]{note_MaierSaupe}%
  \BibitemOpen
  \href@noop {} {}\bibinfo {note} {A simple calculation shows that in 2D, in
  the isotropic phase the Maier-Saupe order parameter
  $S_\mathrm{MS}\equiv\frac{3}{2}\langle
  \cos^2\theta\rangle-\frac{1}{2}=\frac{1}{4}$.}\BibitemShut {Stop}%
\bibitem [{\citenamefont {Vink}(2007)}]{Vink-PRL-2007}%
  \BibitemOpen
  \bibfield  {author} {\bibinfo {author} {\bibfnamefont {R.}~\bibnamefont
  {Vink}},\ }\href@noop {} {\bibfield  {journal} {\bibinfo  {journal} {Phys.
  Rev. Lett.}\ }\textbf {\bibinfo {volume} {98}},\ \bibinfo {pages} {217801}
  (\bibinfo {year} {2007})}\BibitemShut {NoStop}%
\bibitem [{\citenamefont {Jordens}\ \emph {et~al.}(2013)\citenamefont
  {Jordens}, \citenamefont {Isa}, \citenamefont {Usov},\ and\ \citenamefont
  {Mezzenga}}]{Jordens-NatComm-2013}%
  \BibitemOpen
  \bibfield  {author} {\bibinfo {author} {\bibfnamefont {S.}~\bibnamefont
  {Jordens}}, \bibinfo {author} {\bibfnamefont {L.}~\bibnamefont {Isa}},
  \bibinfo {author} {\bibfnamefont {I.}~\bibnamefont {Usov}}, \ and\ \bibinfo
  {author} {\bibfnamefont {R.}~\bibnamefont {Mezzenga}},\ }\href@noop {}
  {\bibfield  {journal} {\bibinfo  {journal} {Nature Commun.}\ }\textbf
  {\bibinfo {volume} {4}},\ \bibinfo {pages} {1917} (\bibinfo {year}
  {2013})}\BibitemShut {NoStop}%
\bibitem [{\citenamefont {Marrucci}\ and\ \citenamefont
  {Maffettone}(1989)}]{Marucci-Macromol-1989}%
  \BibitemOpen
  \bibfield  {author} {\bibinfo {author} {\bibfnamefont {G.}~\bibnamefont
  {Marrucci}}\ and\ \bibinfo {author} {\bibfnamefont {P.~L.}\ \bibnamefont
  {Maffettone}},\ }\href@noop {} {\bibfield  {journal} {\bibinfo  {journal}
  {Macromolecules}\ }\textbf {\bibinfo {volume} {22}},\ \bibinfo {pages} {4076}
  (\bibinfo {year} {1989})}\BibitemShut {NoStop}%
\bibitem [{\citenamefont {Demus}\ and\ \citenamefont
  {Richter}(1978)}]{demus1978}%
  \BibitemOpen
  \bibfield  {author} {\bibinfo {author} {\bibfnamefont {D.}~\bibnamefont
  {Demus}}\ and\ \bibinfo {author} {\bibfnamefont {L.}~\bibnamefont
  {Richter}},\ }\href@noop {} {\emph {\bibinfo {title} {Textures of Liquid
  Crystals}}}\ (\bibinfo  {publisher} {Verlag Chemie Weinheim},\ \bibinfo
  {year} {1978})\BibitemShut {NoStop}%
\bibitem [{\citenamefont {Chaikin}\ and\ \citenamefont
  {Lubensky}(2000)}]{Chaikin}%
  \BibitemOpen
  \bibfield  {author} {\bibinfo {author} {\bibfnamefont {P.~M.}\ \bibnamefont
  {Chaikin}}\ and\ \bibinfo {author} {\bibfnamefont {T.~C.}\ \bibnamefont
  {Lubensky}},\ }\href@noop {} {\emph {\bibinfo {title} {Principles of
  Condensed Matter Physics}}}\ (\bibinfo  {publisher} {Cambridge Univ. Press},\
  \bibinfo {year} {2000})\BibitemShut {NoStop}%
\bibitem [{\citenamefont {Liu}\ \emph {et~al.}(2007)\citenamefont {Liu},
  \citenamefont {Shen},\ and\ \citenamefont {Yang}}]{Liu:2007}%
  \BibitemOpen
  \bibfield  {author} {\bibinfo {author} {\bibfnamefont {C.}~\bibnamefont
  {Liu}}, \bibinfo {author} {\bibfnamefont {J.}~\bibnamefont {Shen}}, \ and\
  \bibinfo {author} {\bibfnamefont {X.}~\bibnamefont {Yang}},\ }\href@noop {}
  {\bibfield  {journal} {\bibinfo  {journal} {Commun. Comput. Phys.}\ }\textbf
  {\bibinfo {volume} {2}},\ \bibinfo {pages} {1184} (\bibinfo {year}
  {2007})}\BibitemShut {NoStop}%
\bibitem [{\citenamefont {{J.-i. Fukuda}}(1998)}]{FukudaEPJB1998}%
  \BibitemOpen
  \bibfield  {author} {\bibinfo {author} {\bibnamefont {{J.-i. Fukuda}}},\
  }\href {\doibase 10.1007/s100510050168} {\bibfield  {journal} {\bibinfo
  {journal} {Eur. Phys. J. B}\ }\textbf {\bibinfo {volume} {1}},\ \bibinfo
  {pages} {173} (\bibinfo {year} {1998})}\BibitemShut {NoStop}%
\bibitem [{\citenamefont {T{\'o}th}\ \emph {et~al.}(2002)\citenamefont
  {T{\'o}th}, \citenamefont {Denniston},\ and\ \citenamefont
  {Yeomans}}]{Yeomans:2002}%
  \BibitemOpen
  \bibfield  {author} {\bibinfo {author} {\bibfnamefont {G.}~\bibnamefont
  {T{\'o}th}}, \bibinfo {author} {\bibfnamefont {C.}~\bibnamefont {Denniston}},
  \ and\ \bibinfo {author} {\bibfnamefont {J.~M.}\ \bibnamefont {Yeomans}},\
  }\href@noop {} {\bibfield  {journal} {\bibinfo  {journal} {Phys. Rev. Lett.}\
  }\textbf {\bibinfo {volume} {88}},\ \bibinfo {pages} {105504} (\bibinfo
  {year} {2002})}\BibitemShut {NoStop}%
\bibitem [{\citenamefont {Sven\ifmmode~\check{s}\else \v{s}\fi{}ek}\ and\
  \citenamefont {\ifmmode~\check{Z}\else
  \v{Z}\fi{}umer}(2002)}]{SvensekPRE2002}%
  \BibitemOpen
  \bibfield  {author} {\bibinfo {author} {\bibfnamefont {D.}~\bibnamefont
  {Sven\ifmmode~\check{s}\else \v{s}\fi{}ek}}\ and\ \bibinfo {author}
  {\bibfnamefont {S.}~\bibnamefont {\ifmmode~\check{Z}\else \v{Z}\fi{}umer}},\
  }\href {\doibase 10.1103/PhysRevE.66.021712} {\bibfield  {journal} {\bibinfo
  {journal} {Phys. Rev. E}\ }\textbf {\bibinfo {volume} {66}},\ \bibinfo
  {pages} {021712} (\bibinfo {year} {2002})}\BibitemShut {NoStop}%
\bibitem [{\citenamefont {Blanc}\ \emph {et~al.}(2005)\citenamefont {Blanc},
  \citenamefont {Sven\ifmmode~\check{s}\else \v{s}\fi{}ek}, \citenamefont
  {\ifmmode~\check{Z}\else \v{Z}\fi{}umer},\ and\ \citenamefont
  {Nobili}}]{BlancPRL2005}%
  \BibitemOpen
  \bibfield  {author} {\bibinfo {author} {\bibfnamefont {C.}~\bibnamefont
  {Blanc}}, \bibinfo {author} {\bibfnamefont {D.}~\bibnamefont
  {Sven\ifmmode~\check{s}\else \v{s}\fi{}ek}}, \bibinfo {author} {\bibfnamefont
  {S.}~\bibnamefont {\ifmmode~\check{Z}\else \v{Z}\fi{}umer}}, \ and\ \bibinfo
  {author} {\bibfnamefont {M.}~\bibnamefont {Nobili}},\ }\href {\doibase
  10.1103/PhysRevLett.95.097802} {\bibfield  {journal} {\bibinfo  {journal}
  {Phys. Rev. Lett.}\ }\textbf {\bibinfo {volume} {95}},\ \bibinfo {pages}
  {097802} (\bibinfo {year} {2005})}\BibitemShut {NoStop}%
\bibitem [{\citenamefont {Denniston}(1996)}]{Denniston:1996}%
  \BibitemOpen
  \bibfield  {author} {\bibinfo {author} {\bibfnamefont {C.}~\bibnamefont
  {Denniston}},\ }\href@noop {} {\bibfield  {journal} {\bibinfo  {journal}
  {Phys. Rev. B}\ }\textbf {\bibinfo {volume} {54}},\ \bibinfo {pages} {6272}
  (\bibinfo {year} {1996})}\BibitemShut {NoStop}%
\bibitem [{\citenamefont {Mather}\ \emph {et~al.}(1997)\citenamefont {Mather},
  \citenamefont {Romo-Uribe}, \citenamefont {Han},\ and\ \citenamefont
  {Kim}}]{Mather-Macromol-1997}%
  \BibitemOpen
  \bibfield  {author} {\bibinfo {author} {\bibfnamefont {P.~T.}\ \bibnamefont
  {Mather}}, \bibinfo {author} {\bibfnamefont {A.}~\bibnamefont {Romo-Uribe}},
  \bibinfo {author} {\bibfnamefont {C.~D.}\ \bibnamefont {Han}}, \ and\
  \bibinfo {author} {\bibfnamefont {S.~S.}\ \bibnamefont {Kim}},\ }\href
  {\doibase 10.1021/ma970737h} {\bibfield  {journal} {\bibinfo  {journal}
  {Macromolecules}\ }\textbf {\bibinfo {volume} {30}},\ \bibinfo {pages} {7977}
  (\bibinfo {year} {1997})}\BibitemShut {NoStop}%
\bibitem [{\citenamefont {Olmsted}(2008)}]{Olmsted-Rheo-2008}%
  \BibitemOpen
  \bibfield  {author} {\bibinfo {author} {\bibfnamefont {P.}~\bibnamefont
  {Olmsted}},\ }\href {\doibase 10.1007/s00397-008-0260-9} {\bibfield
  {journal} {\bibinfo  {journal} {Rheologica Acta}\ }\textbf {\bibinfo {volume}
  {47}},\ \bibinfo {pages} {283} (\bibinfo {year} {2008})}\BibitemShut
  {NoStop}%
\end{thebibliography}
%

\end{document}